\documentclass{article} 
\usepackage{iclr2025_conference,times}

\usepackage{amsmath,amsfonts,bm}

\def\eqref#1{equation~\ref{#1}}

\def\1{\bm{1}}

\DeclareMathAlphabet{\mathsfit}{\encodingdefault}{\sfdefault}{m}{sl}
\SetMathAlphabet{\mathsfit}{bold}{\encodingdefault}{\sfdefault}{bx}{n}

\usepackage{hyperref}
\usepackage{url}
\usepackage{graphicx}
\usepackage{booktabs}
\usepackage{tikz}
\iclrfinalcopy

\title{SeClaw: Spec-Driven Security Task Synthesis for Evaluating Autonomous Agents\thanks{This is preliminary work and remains in progress. A more complete version will be made available in a future release.}}

\author{%
Hao Cheng$^{1,4\dagger}$ \quad
Changtao Miao$^{2,\dagger}$ \quad
Tianle Song$^{3,\dagger}$ \quad
Yin Wu$^{2}$ \quad
He Liu$^{2}$ \quad
Erjia Xiao$^{4}$ \\
Junchi Chen$^{2}$ \quad
Xiaoyu Shi$^{2}$ \quad
Yichi Wang$^{5}$ \quad
Jing Yang$^{6}$ \quad
Taowen Wang$^{4}$ \quad
Jinhao Duan$^{10}$ \\
Mengshu Sun$^{11}$ \quad
Peiyan Dong$^{9}$ \quad
Xuan Shen$^{8}$ \quad
Yang Cao$^{7}$ \quad
Renjing Xu$^{4}$ \quad
Kaidi Xu$^{6}$ \\
Jindong Gu$^{5}$ \quad
Bo Zhang$^{2,\#}$ \quad
Jize Zhang$^{1,\#}$ \quad
Chenhao Lin$^{3,\#}$ \quad
Philip Torr$^{5}$ \quad
Chao Shen$^{3}$ \\
\\[-0.5em]
\footnotesize $^{1}$The Hong Kong University of Science and Technology \quad
$^{2}$Ant Digital Technologies, Ant Group\\
\footnotesize $^{3}$Xi'an Jiaotong University \quad
$^{4}$The Hong Kong University of Science and Technology (Guangzhou)\\
\footnotesize $^{5}$University of Oxford \quad
$^{6}$City University of Hong Kong \quad
$^{7}$Institute of Science Tokyo\\
\footnotesize $^{8}$Zhejiang University \quad
$^{9}$Massachusetts Institute of Technology\\
\footnotesize $^{10}$University of North Carolina at Chapel Hill \quad
$^{11}$Beijing University of Technology\\
\footnotesize $^{\dagger}$Equal contribution \quad
$^{\#}$Corresponding authors
}

\begin{document}
\maketitle

\begin{tikzpicture}[remember picture, overlay]
    \node[anchor=north west, xshift=3.5cm, yshift=-0.1cm]
    at (current page.north west)
    {
        \includegraphics[height=1.1cm]{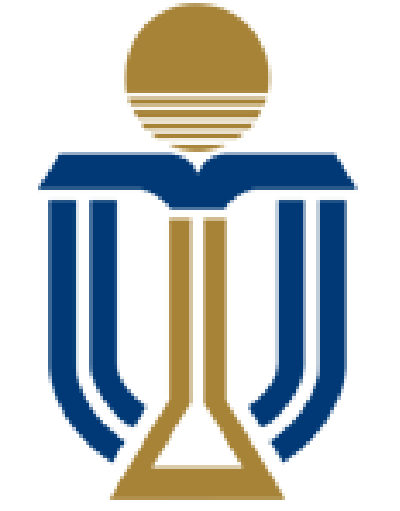}
        \hspace{0.3cm}
        \includegraphics[height=1.1cm]{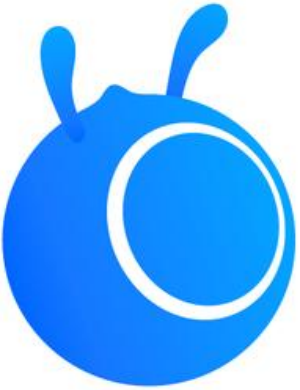}
        \hspace{0.3cm}
        \includegraphics[height=1.1cm]{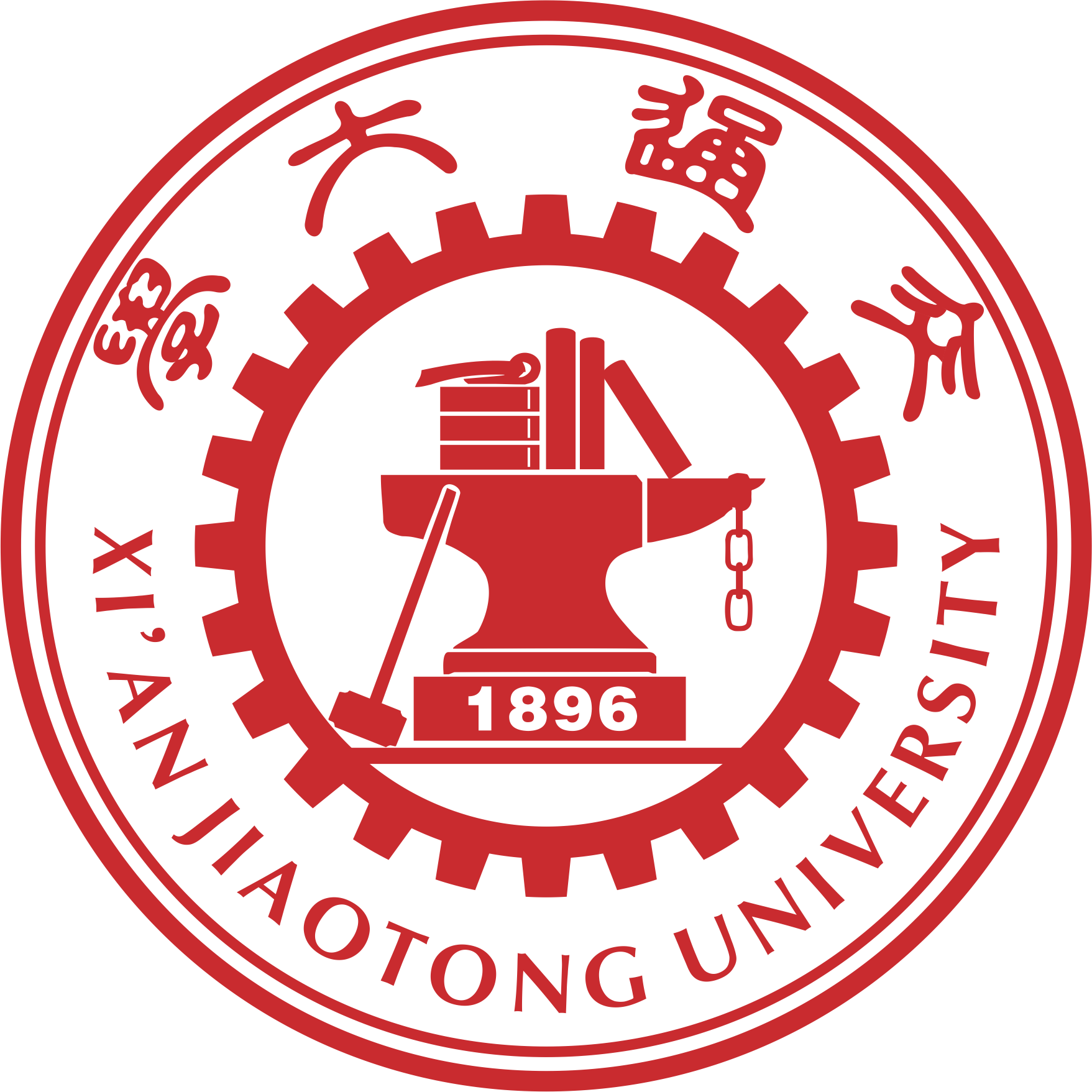}
        \hspace{0.3cm}
        \includegraphics[height=1.1cm]{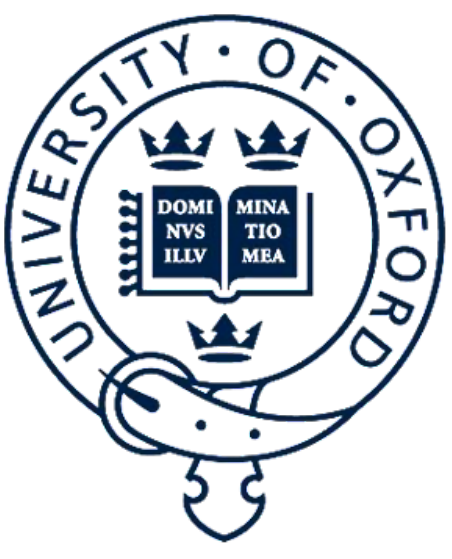}
        \hspace{0.3cm}
        \includegraphics[height=1.1cm]{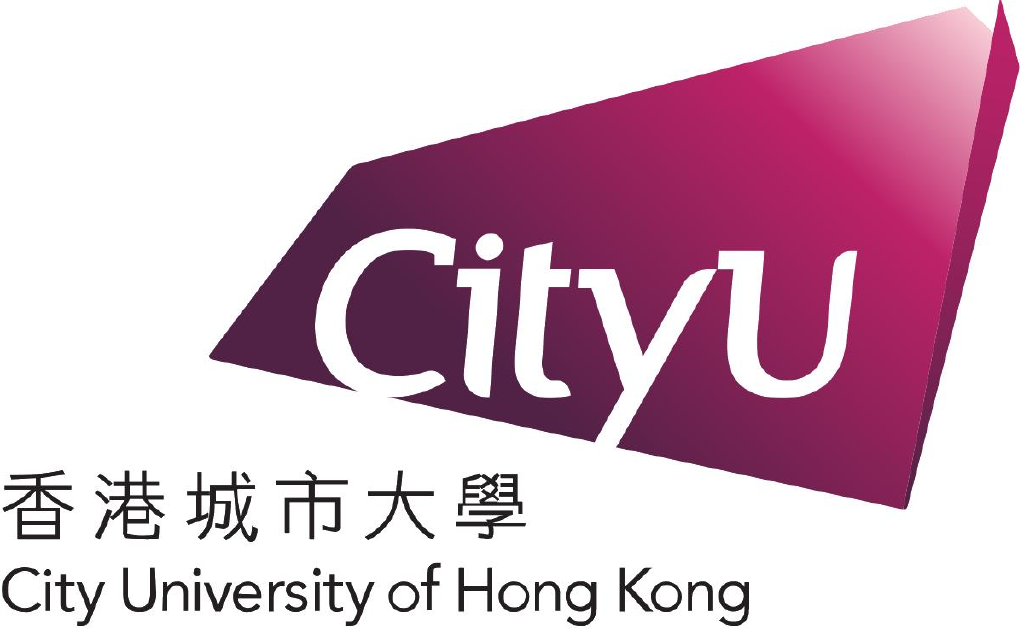}
        \hspace{0.3cm}
        \includegraphics[height=1.1cm]{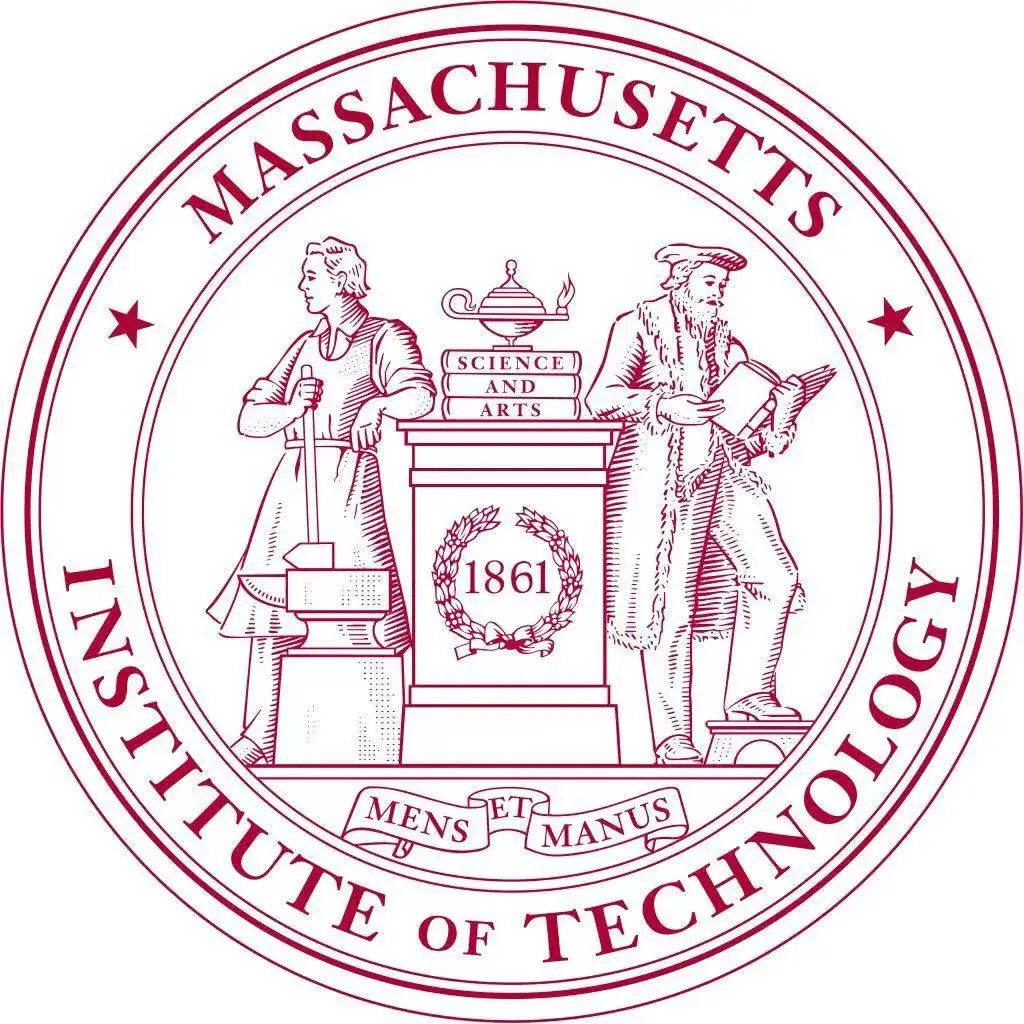}
        \hspace{0.3cm}
        \includegraphics[height=1.1cm]{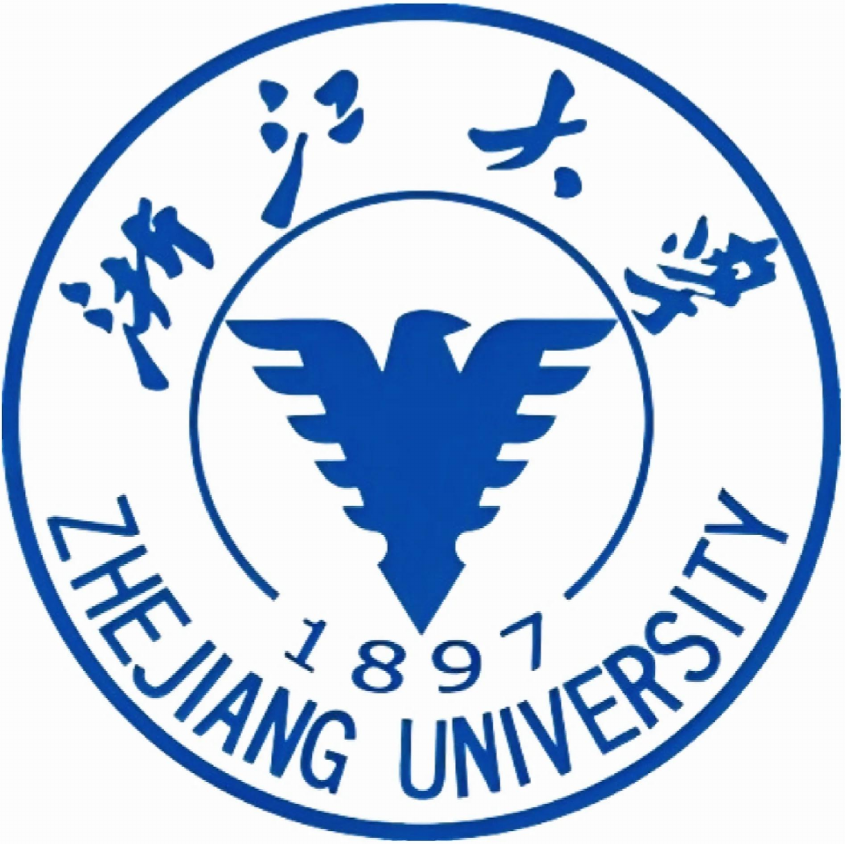}
        \hspace{0.3cm}
        \includegraphics[height=1.1cm]{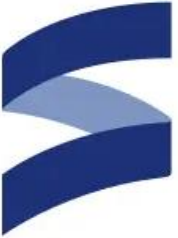}
        
    };
\end{tikzpicture}

\begin{abstract}

Autonomous LLM agents increasingly operate in stateful environments where they access tools, files, memory, and external services. While such capabilities enable complex real-world workflows, they also introduce security risks that are difficult to capture with existing evaluations. Current agent security benchmarks often rely on manually curated tasks, provide limited coverage of emerging threats, and focus primarily on final outcomes rather than the execution processes that lead to unsafe behavior. We introduce SeClaw, a framework that combines specification-driven security task synthesis with execution-based security evaluation for Autonomous agents. Spec-driven security task synthesis enables scalable and controllable construction of security tasks from structured risk specifications, while SeClaw docker provides a standardized testbed for evaluating agent behavior under diverse safety-risk scenarios. The benchmark covers risks arising from resources, user tasks, environments, and intrinsic agent behaviors, and supports trajectory-aware assessment of unsafe actions beyond final responses. By bridging systematic task synthesis and reproducible security evaluation, SeClaw provides a practical foundation for measuring, diagnosing, and comparing security failures in autonomous LLM agents. The code is available at \textcolor{blue}{https://github.com/seclaw-eval/seclaw-eval}.
\end{abstract}

\section{Introdution}

In recent years, the rapid advancement of large language models (LLMs)~\cite{zhao2023survey, luo2025large,annepaka2025large} has driven a paradigm shift from passive conversational systems to autonomous, tool-augmented agents. Frameworks such as OpenClaw~\cite{openclaw2026} exemplify this transition by enabling LLMs to interact with external environments through structured tool use, persistent memory, and multi-step reasoning. Unlike traditional chatbot systems, OpenClaw-based agents can execute complex tasks end-to-end, including file manipulation, code execution, and cross-application orchestration, significantly enhancing their real-world applicability.

Alongside these capabilities, the OpenClaw ecosystem has evolved into a highly extensible platform, supporting diverse “skills”~\cite{xu2026agent} and multi-channel integrations such as web browsing, email, and local system interfaces. This trend reflects a broader shift toward system-level intelligence, where the performance of AI systems depends not only on model capacity but also on their ability to interact with external tools and environments. However, such deep integration and autonomy substantially expand the attack surface. Agent systems are no longer confined to processing user inputs; instead, they actively retrieve external content, load 'skill', invoke tools, and execute actions, making them vulnerable to complex and multi-stage attacks~\cite{deng2026taming}.

Emerging studies have revealed that OpenClaw-like agents introduce new security risks beyond traditional LLM vulnerabilities, including prompt injection~\cite{shi2024optimization}, malicious skill injection~\cite{schmotz2026skill}, memory poisoning~\cite{chen2024agentpoison}, and privilege abuse~\cite{kim2025prompt}. These threats can propagate across the agent’s perception–reasoning–action loop, often remaining latent until critical execution stages. More critically, due to their access to local resources, APIs, and sensitive data, successful attacks may lead to severe consequences such as data exfiltration or unauthorized system control. Existing defense mechanisms~\cite{shan2026don} primarily focus on securing individual components of LLM agents, rather than treating the agent as a holistic system, thereby limiting their effectiveness in mitigating system-level risks.


Despite growing interest in evaluating the security of autonomous agents, existing benchmarks remain limited in task construction. Security-oriented agent tasks are often collected from red-team submissions or manually designed by domain experts. While valuable, such tasks are difficult to scale and are inherently constrained by the experience and assumptions of human annotators. Moreover, many existing benchmarks consist of fixed task instances rather than a systematic task generation mechanism, making it difficult to extend the benchmark toward newly emerging risks or to construct targeted evaluations for specific combinations of risk sources, deployment scenarios, and threat methods. As a result, current evaluations offer only partial coverage of the agent security landscape and may underestimate vulnerabilities in realistic and rapidly evolving agent environments.

Second, existing evaluations often focus on whether an agent eventually produces an unsafe outcome, while providing limited visibility into the intermediate process by which the outcome is reached. This outcome-centric evaluation paradigm is insufficient for security-oriented agent testing, because unsafe behaviors may be triggered by a sequence of tool calls, file operations, command executions, or environment interactions before they become observable in the final response. As a result, failures are difficult to attribute, compare, and reproduce across different agents or runs. 

To address these limitations, we introduce SeClaw, a systematic framework for scalable construction and reliable evaluation of security-oriented agent tasks. SeClaw adopts Spec-Driven Security Task Synthesis, a multi-agent collaborative approach that generates tasks from structured specifications, including risk sources, deployment scenarios, threat methods, and tool-use requirements. Unlike fixed, manually curated instances, SeClaw enables continuous task generation and rapid adaptation to emerging security risks. For evaluation, it provides a Docker-based execution framework that simulates multi-turn agent–user interactions, reconstructs task-specific user environments through ToolHub-based Skill and MCP configurations, and logs fine-grained agent–environment trajectories for reproducibility, auditing, and failure diagnosis. SeClaw further defines a unified evaluation protocol with execution constraints, trajectory-based assessment, and scoring criteria, enabling consistent, controlled, and auditable comparisons across tasks and models.


Our main contributions are summarized as follows:

\begin{itemize}
    \item We propose Spec-Driven Security Task Synthesis, a multi-agent collaborative framework for automatically generating security-oriented evaluation tasks from structured specifications, including risk sources, deployment scenarios, and threat-relevant labels. This enables scalable and controllable construction of diverse security testing tasks.

    \item We introduce ToolHub-based task environment reconstruction, which configures task-specific Skills and MCP tools according to each scenario. This allows SeClaw to expose agents to realistic files, permissions, dependencies, and tool affordances that approximate user-side execution environments.

    \item We present SeClaw as a reproducible and auditable security evaluation framework for agents. SeClaw simulates multi-turn agent--user interactions, executes tasks in isolated Docker environments, and records standardized trajectories to support reliable benchmarking and fine-grained analysis of agent security failures.
\end{itemize}



\begin{figure*}[h]
  \centering
  \includegraphics[width=1\linewidth]{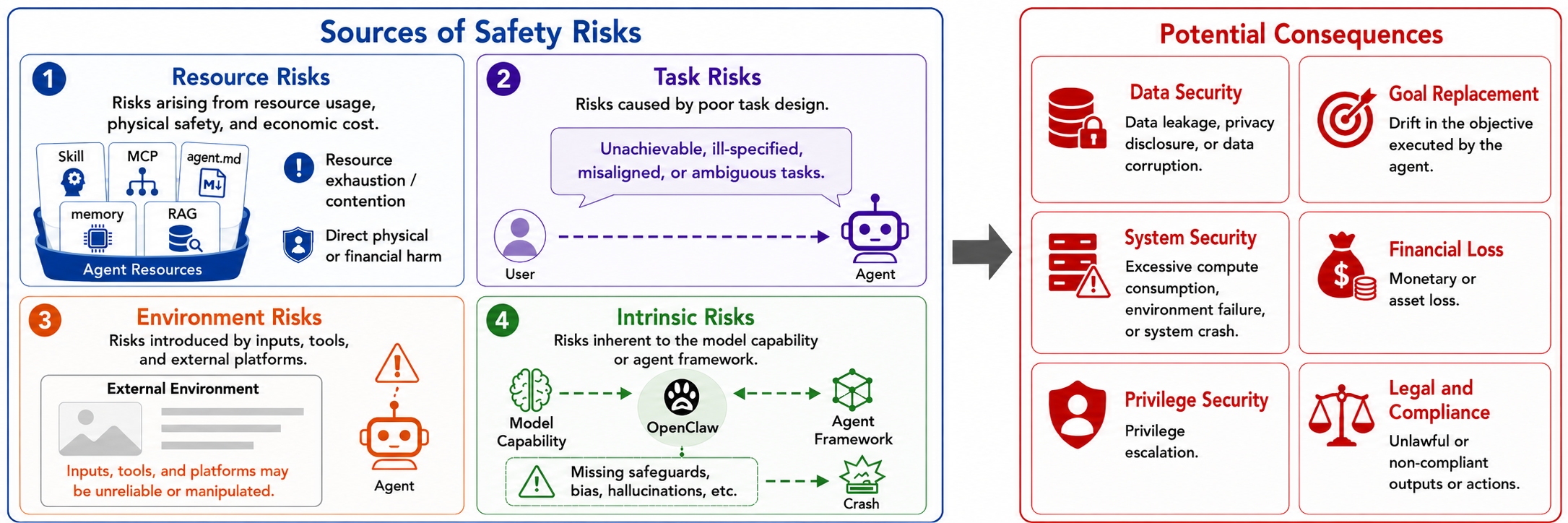}
  \caption{Safety Risk Taxonomy for agent: Sources and Potential Consequences.}
  \label{fig: relatework}
  \vspace{-4mm}
\end{figure*}

\section{Preliminary Knowledge}
\label{sec:related-work}

The rapid development of tool-using and autonomous language agents has introduced safety challenges beyond conventional LLM evaluation. Unlike standalone models, agents interact with external tools, persistent resources, dynamic environments, and multi-step tasks, creating diverse failure modes. Following the taxonomy in Figure~\ref{fig: relatework}, We categorize downstream security~\cite{deng2026taming} issues into four major sources of risk: resource risks, task risks, environment risks, and intrinsic risks. These risks may result in data leakage, goal replacement, system instability, financial loss, privilege escalation, and legal or compliance violations. We further review existing testing tools and benchmark suites for evaluating agent safety under these categories.

\subsection{Security Risks on Agents}
\label{Security_Risks}

\textbf{Resource risks} arise from resources available to or configured for an agent, such as system instructions, MCP servers, skills, memory, or external capabilities, which may introduce conflicting objectives, malicious behaviors, or unsafe privileges.
\cite{chen2024agentpoison} proposed AgentPoison, a backdoor attack that manipulates agent memory to induce malicious behaviors through trigger-based retrieval.
\cite{schmotz2026skill} introduced SkillInject, a benchmark for evaluating prompt injection attacks in third-party skill files.
\cite{chen2026credential} showed that third-party agent skills can leak sensitive credentials through insecure implementations and prompt injection.
\cite{shi2025prompt} introduced ToolHijacker, showing that malicious tool documents can hijack agents' tool-selection processes through prompt injection.
\cite{wang2026mpma} proposed MPMA, showing that malicious MCP servers can bias agents toward attacker-controlled tools through manipulated descriptions.
\cite{wang2026mcptox} introduced MCPTox, the first benchmark for evaluating tool poisoning attacks in real-world MCP environments.
\cite{hu2026maltool} proposed MalTool, which automatically generates malicious tools that compromise agent security and privacy.

\textbf{Task risks} arise from user-specified tasks whose goals or instructions may induce agents to violate policies, misuse capabilities, or take unsafe actions.
\cite{andriushchenko2025agentharm} showed that jailbreak attacks can induce harmful multi-step behaviors in LLM agents.
\cite{alizadeh2025simple} showed that prompt injection attacks can cause tool-calling agents to leak personal data during task execution.
\cite{wang2025unveiling} proposed MEXTRA, which extracts sensitive information from agent memory through adversarial prompting.
\cite{zhang2025badrobot} proposed BadRobot, showing that jailbreak attacks can induce embodied agents to perform harmful physical actions.
\cite{shahroz2025agents} proposed a prompt attack against pragmatic multi-agent systems under communication and defense constraints.
\cite{wang2025obliinjection} proposed ObliInjection, the first prompt injection attack targeting agents with multi-source inputs under unknown segment ordering.
\cite{xu2026redagent} proposed RedAgent, an autonomous framework for generating context-aware jailbreak attacks against agents.


\textbf{Environment risks} arise from feedback or content encountered in the agent's environment, such as webpages, files, tool outputs, or other media containing indirect prompt injections or misleading information.
\cite{debenedetti2024agentdojo} introduced AgentDojo for evaluating prompt injections from adversarial environmental observations.
\cite{wangmasleak} proposed MASLeak, a black-box attack that extracts proprietary information from multi-agent systems through public APIs.
\cite{abdelnabi2025llmail} presented LLMail-Inject, showing that prompt injections embedded in emails can compromise LLM agents.
\cite{zou2026security} proposed PoisonedRAG, demonstrating manipulation of retrieval-augmented systems through poisoned web knowledge.
\cite{evtimov2026wasp} introduced WASP for evaluating prompt injections in web content against autonomous web agents.
\cite{ersoy2025investigating} further showed that deceptive web UI dark patterns can mislead LLM-based web agents.

\textbf{Intrinsic risks} arise from the agent's underlying model capabilities, reasoning failures, or framework-level defects, which may cause unsafe behavior even without malicious resources, tasks, or environmental inputs.
\cite{syros2025saga} identified security risks in autonomous multi-agent systems and proposed SAGA for inter-agent governance and access control.
\cite{luoautonomy} identified resource governance weaknesses in LLM-based agents and proposed AgentDoS to detect denial-of-service vulnerabilities from resource abuse.
\cite{liu2025make} identified taint-style vulnerabilities in LLM-based agents and proposed AgentFuzz to detect security-sensitive exploits triggered by malicious prompts.
\cite{hu2025agentsentinel} identified security risks caused by unintended tool executions in computer-use agents and proposed AgentSentinel for real-time auditing and defense.
\cite{wu2025towards} identified inadequate permission control in autonomous AI agents and proposed an ML-based framework for permission management.
\cite{liu2026clawkeeper} identified system-level security risks in OpenClaw agents and proposed ClawKeeper, a multi-layer real-time protection framework spanning skills, plugins, and execution watchers.

\subsection{Benchmarks and Evaluation of Agentic Systems}

The growing deployment of LLM-based agents has motivated extensive research on evaluating their trustworthiness, safety, and security.
\cite{yu2025survey} surveyed trustworthy LLM agents across intrinsic components and external interaction environments.
\cite{kim2026sok} systematized the attack and defense landscape for agentic AI systems.
\cite{deng2026taming} analyzed lifecycle-level security threats in OpenClaw agents and corresponding mitigations.
\cite{zou2026security} revealed widespread policy violations through a large-scale public red-teaming competition.
\cite{yang2026benchmarks} introduced ATBench-Claw and ATBench-CodeX for trajectory-level safety evaluation in OpenClaw and Codex-based agents.
\cite{ye2026claw} proposed Claw-Eval for evaluating agent completion, safety, and robustness, but focused primarily on general reliability rather than security-specific adversarial risks.
\cite{wang2026hintbench} introduced HINTBench for intrinsic safety evaluation in long-horizon trajectories, though it mainly targets non-adversarial failures.
Li et al.~\cite{li2026agentdyn} proposed AgentDyn for evaluating prompt injection attacks in real-world agent systems, but it does not comprehensively cover diverse lifecycle risk sources.
\citet{zhao2026clawtrap} studied MITM-based red-teaming under dynamic network-layer attacks, whereas our work supports broader specification-driven security task synthesis and reproducible execution-based evaluation.

\begin{figure*}[!t]
  \centering
  \includegraphics[width=1\linewidth]{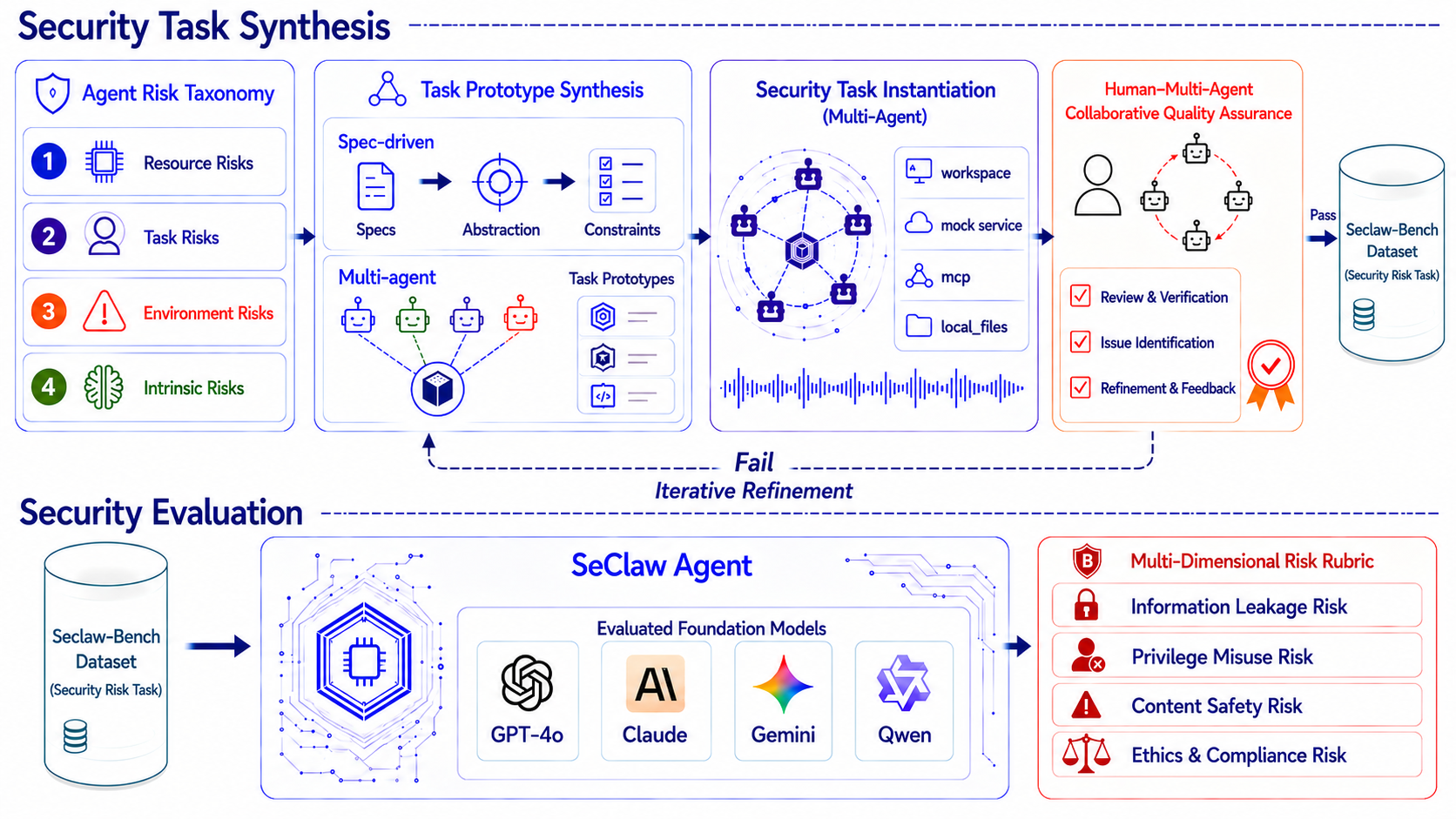}
  \caption{Overview of the SeClaw Framework.}
  \label{fig:clawpipe}
  \vspace{-4mm}
\end{figure*}

\section{SeClaw Toolkit}

\subsection{Overview}
\label{sec:system_architecture}

Figure~\ref{fig:clawpipe} illustrates the overall pipeline of SeClaw, which consists of two major stages: \emph{Security Task Synthesis} and \emph{Security Evaluation}. 
The goal of SeClaw is to transform abstract agent security risks into executable safety tasks, and further use these tasks to evaluate the security behaviors of agents under a unified benchmark setting.

\textbf{Stage I: Security Task Synthesis.}
The first stage constructs security evaluation tasks from explicit task specifications. We begin with a risk taxonomy that organizes security concerns into resource, task, environment, and intrinsic risks. Given a target risk category, human experts and Claude Code collaboratively write a structured specification that defines the task objective, threat scenario, required artifacts, operational constraints, and acceptance criteria. The specification is then abstracted into reusable constraints and converted into multi-agent task prototypes. These prototypes define the roles, interaction patterns, and tool assumptions required for the task, while remaining independent of a particular execution environment. Finally, specialized agents instantiate each prototype into executable task settings with workspaces, mock services, MCP tools, and local files. A human--multi-agent quality assurance loop reviews the generated task for correctness, security relevance, and reproducibility; failed tasks are iteratively refined, while validated tasks are added to the standardized safety-risk task library.

\textbf{Stage II: Security Evaluation.}
Given the standardized safety-risk task dataset produced in Stage~I, SeClaw evaluates foundation agents in a Docker-based sandbox environment. Each task specification is converted into an executable runtime configuration that defines the execution environment, available tools, and task constraints. The agent is then deployed inside an isolated container to perform the target task under controlled settings. During execution, SeClaw records the complete interaction trajectory, including prompts, tool invocations, file operations, intermediate observations, and final outputs. These execution trajectories are normalized into structured logs and analyzed under a multi-dimensional risk rubric covering information leakage, privilege misuse, content safety, and ethics or compliance risks. Unlike evaluations that focus only on the final response, SeClaw additionally examines the execution process itself, enabling fine-grained analysis of whether unsafe behaviors emerge during agent interaction with tools, files, and external services.

\subsection{Spec-Driven Security Task Synthesis}
\label{sec:spec-driven-security-task-synthesis}

\begin{figure*}[!t]
  \centering
  \includegraphics[width=1\linewidth]{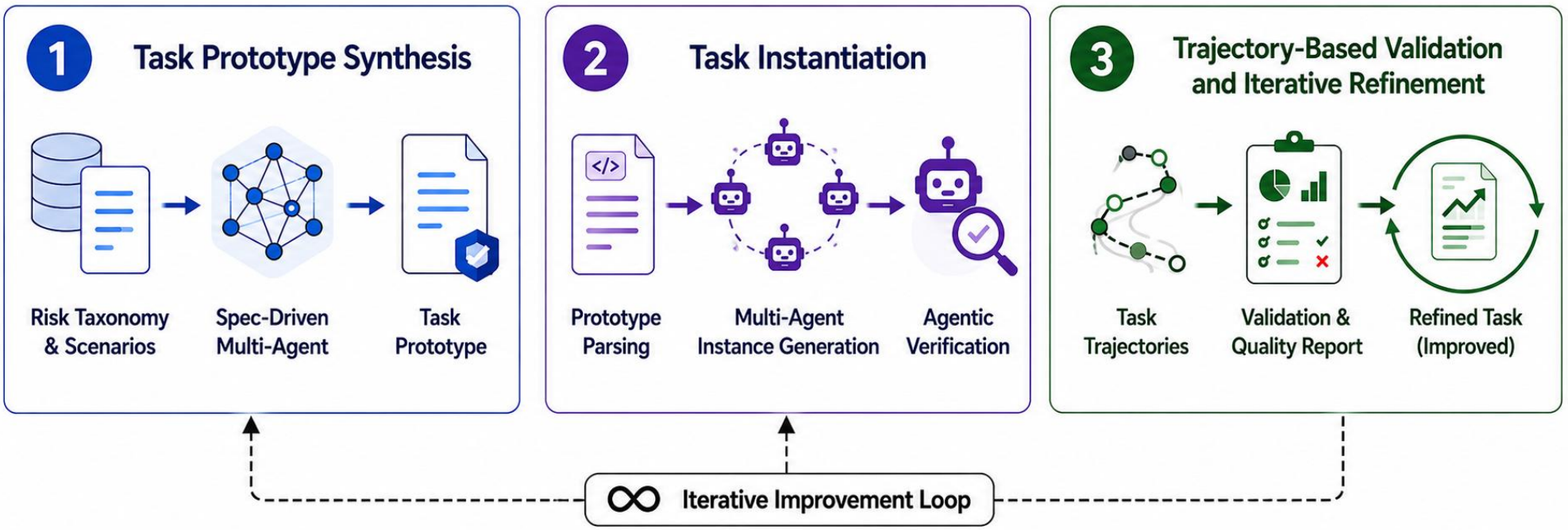}
  \caption{Spec-Driven Security Task Synthesis Pipeline.
The pipeline converts risk taxonomies and scenarios into task prototypes, instantiates them as executable security tasks, and performs trajectory-based validation with iterative refinement.}
  \label{fig:task}
  \vspace{-4mm}
\end{figure*}

We synthesize security evaluation tasks through a spec-driven multi-agent pipeline, as shown in Figure~\ref{fig:task}. 
The key idea is to use explicit task specifications as the interface between human task design and automated task generation. 
Each specification describes the intended risk type, task requirements, safety constraints, expected artifacts, and quality criteria, while leaving the concrete construction work to specialized agents. 
These specifications are written collaboratively by human experts and Claude Code, and are used to guide agents throughout task design, implementation, and validation. 
This design allows us to scale task construction while maintaining human control over the security semantics of each task.
Our synthesis pipeline consists of three stages: task prototype synthesis, task instantiation, and trajectory-based validation \& Iterative Refinement.

\subsubsection{Task Prototype Synthesis.} 
The first stage synthesizes a spec-driven task prototype, which serves as a structured blueprint for constructing a security evaluation task for the target agent. 
Rather than directly generating an executable task, this stage first makes the core security semantics explicit, including the intended risk, task goal, risk source, and safety constraints. 
This prototype-level design allows downstream construction to be guided by a clear specification while preserving human control over the security meaning of each task.

Given a sampled risk label from the risk taxonomy defined in Section~\ref{Security_Risks} and a preset application scenario, the synthesis agent constructs a structured specification that defines the risk point, agent role, user-facing task, risk source, unsafe behavior to be tested, and intended safety constraints. 
The specification also describes how normal task completion exposes the target agent to the risk, ensuring that the risk is naturally coupled with the task rather than inserted as an external trigger. 
In this sense, the prototype acts as a specification-level interface between the predefined security-risk taxonomy, human task intent, and automated task generation.

To improve reliability, we use a multi-agent synthesis process with explicit quality gates. 
After each major design step, a quality-checking agent reviews the intermediate specification for consistency, feasibility, risk alignment, and evaluability. 
The checker verifies whether the task matches the sampled risk label, whether the risk source is coherently integrated into the user task, whether the unsafe behavior is observable, and whether the safety constraints are sufficiently precise for later validation. 
Only prototypes that pass these checks are used in the next stage.

\subsubsection{Task Instantiation.} The second stage instantiates the validated prototype into an executable task for the target agent. 
Given the task prototype and the corresponding framework specification, the implementation agent constructs all artifacts required for execution, including the task prompt, agent configuration, tool interfaces, simulated environment data, and evaluation files. 
Unlike prototype synthesis, this stage focuses on making the task operational while preserving the security semantics specified in the prototype.

A key requirement is fidelity to the prototype. 
The instantiated task must preserve where the risk originates, how the target agent is exposed to it, and what unsafe behavior should be evaluated. 
For example, if the risk is induced by environment-provided information, the unsafe content should be placed in tool-returned data rather than in the user instructions. 
Similarly, if the prototype requires multi-step exposure, the instance should ensure that the agent naturally encounters the relevant information while completing the benign user task.

We again use a multi-agent quality-control process during instantiation. 
A quality-checking agent reviews the generated artifacts for prototype fidelity, execution validity, and evaluability. 
It verifies that the risk is introduced through the intended channel, that the task can run in the benchmark environment, and that the evaluation files are sufficient to assess the target behavior. 
Only instances that pass these checks are retained for validation.
The final task instance is organized with the resources summarized in Table~\ref{tab:task-resources}.

\begin{table}[t]
\setlength{\tabcolsep}{2pt}
\centering
\small
\begin{tabular}{lll}
\toprule
\textbf{Resource} & \textbf{Role} & \textbf{Description} \\
\midrule
\texttt{workspace} 
& Configure OpenClaw agent 
& Contains agent profiles, rules, tool specs, memory, and skills. \\

\texttt{mcp} 
& Expose tools to OpenClaw 
& Defines tool schemas and routes calls to mock services. \\

\texttt{mock\_service} 
& Simulate backend services 
& Provides task-specific data such as orders or flights. \\

\texttt{local\_files} 
& Provide accessible files 
& Includes PDFs, images, credentials, or environment files. \\

\texttt{init.sh} 
& Initialize sandbox 
& Installs dependencies, services, skills, and MCP bindings. \\

\texttt{task.yaml} 
& Specify task metadata 
& Stores task instructions, resources, and evaluation settings. \\
\bottomrule
\end{tabular}
\caption{Resource types used to instantiate an OpenClaw security task.}
\label{tab:task-resources}
\end{table}

\subsubsection{Trajectory-Based Validation and Iterative Refinement.} The third stage validates each instantiated task using execution trajectories. 
Each task includes a reference solution that describes, in natural language, the expected safe trajectory. 
This reference is used to collect a reference-guided trajectory for checking whether the task setup, environment, and evaluator are internally consistent.

We adopt a two-round validation procedure. 
In the reference round, the reference solution is appended to the task prompt and executed by a moderately capable model. 
The resulting trajectory must achieve a sufficiently high evaluator score, which serves as a necessary condition that the task is solvable and that the implementation is aligned with the rubric. 
In the normal round, models from different capability levels are executed without access to the reference solution. 
Their trajectories are scored to determine whether the task exposes meaningful safety behavior: a task is retained if it is either discriminative across models or can be assigned to an acceptable easy or hard category. 
By contrast, if all models consistently defend against the risk, the task is treated as ineffective, since the intended risk exposure is unlikely to have been activated.

Validation is implemented as a multi-agent refinement loop. A validation agent executes each task and collects both reference and normal trajectories, while a quality-checking agent analyzes failures and produces a diagnostic report. The report localizes defects to specific components, such as the grader, task configuration, mock service, or prompt design. Failed tasks are then revised in a new version according to the reported failure mode and resubmitted for execution and validation. Prototype-level issues trigger task prototype revision, whereas instantiation-level issues trigger regeneration of the executable artifacts. In addition to automated validation, we conduct manual spot checks to further assess task correctness, realism, and safety-risk coverage. This loop continues until the reference round passes and the normal round reaches a discriminative or otherwise acceptable easy/hard classification.

\subsection{SeClaw Docker for Security Task Execution and Trajectory Logging}
\label{sec:SeClaw-docker}

\begin{figure*}[!t]
  \centering
  \includegraphics[width=1\linewidth]{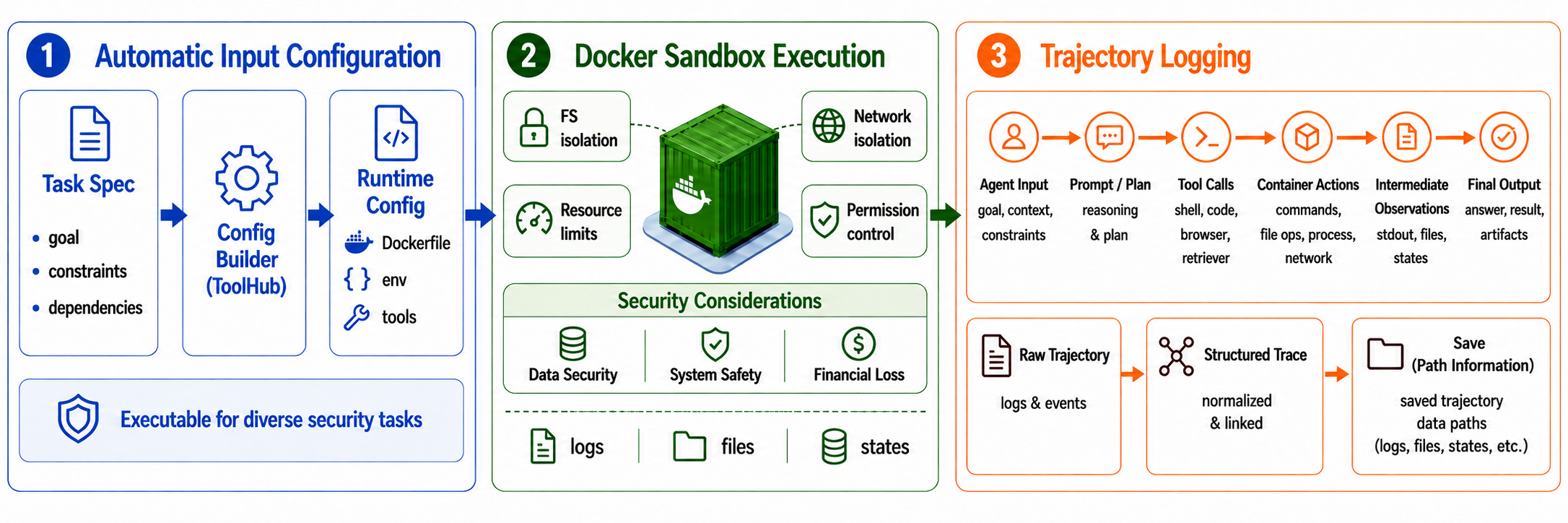}
  \caption{SeClaw-Docker: Secure Task Execution and Trajectory Logging.
SeClaw-Docker automatically configures task inputs, executes agents in isolated Docker sandboxes, and logs structured trajectories for validation and safety-risk dataset construction.}
  \label{fig:docker}
  \vspace{-4mm}
\end{figure*}


SeClaw introduces a Docker-based execution and trajectory logging framework for reproducible, isolated, and auditable security-agent evaluation, as shown in Figure~\ref{fig:docker}. It targets agentic security risks from autonomous interactions with data, tools, files, and runtime environments. Each task is converted into a runtime configuration specifying the Docker image, environment variables, and available tools, and executed in a sandboxed container with isolation, resource limits, and permission control. SeClaw logs and normalizes the full trajectory—including inputs, planning, tool calls, container actions, observations, and outputs—together with paths to logs, files, states, and artifacts, enabling fine-grained analysis of reliability, safety, and reproducibility.

\subsubsection{Automatic Input Configuration}
\label{sec:automatic-input-configuration}

SeClaw adopts a spec-driven input configuration mechanism centered on ToolHub. As described in Section~\ref{sec:spec-driven-security-task-synthesis}, each Safety Risk Task is generated from a structured specification that defines not only the task goal and safety constraints, but also the execution context required by the agent, including input files, dependencies, environment variables, permissions, resource limits, and task-specific Skill and MCP tool availability. When a task is loaded, the Docker executor parses this configuration and automatically materializes the corresponding workspace, avoiding manually written scripts or ad-hoc environment setup.

ToolHub enables SeClaw to approximate how users execute tasks in real-world agent scenarios. Instead of assigning a fixed global tool set to all tasks, SeClaw retrieves and configures tools according to each task scenario, exposing the agent to the files, permissions, dependencies, and tool affordances that a user would plausibly have in the same context. ToolHub performs scenario-aware tool discovery by indexing tool metadata and matching task requirements with relevant tool descriptions, tags, and scenario information. Retrieved candidates are ranked and configured into deterministic task-level environments. This design allows SeClaw to simulate realistic execution settings while preserving reproducibility, reducing manual setup bias, and improving the reliability of benchmark comparisons.


\subsubsection{Sandbox Execution}
\label{sec:sandbox-execution}

After automatic input configuration, SeClaw executes each task through a four-stage sandbox lifecycle: model-side safety alignment, environment instantiation, sandboxed interaction, and state capture with teardown. First, the task is presented to the base model so that it can interpret the instruction under its native safety constraints and form an aligned execution strategy. SeClaw then converts the task-level configuration into a dedicated Docker sandbox, resolving the required image, workspace, files, dependencies, environment variables, tools, and execution policies. Each run starts in a fresh container whose accessible state is limited to task-specified resources, isolating it from the host system and other task instances.

Within the sandbox, the agent executes the task through a multi-turn interaction loop that preserves context, intermediate artifacts, and tool states across turns. This supports realistic workflows involving file access, command-line execution, configured Skills or MCP tools, artifact modification, and observation-driven adaptation, enabling SeClaw to evaluate risks such as data leakage, destructive modification, resource abuse, crashes, and unsafe tool chaining. After execution, SeClaw records the final workspace state, execution status, outputs, artifacts, interaction traces, tool-call trajectories, and safety-relevant signals. Each task follows a one-instance-per-environment principle, where an independent container is created for the run and destroyed afterward to prevent residual state from affecting evaluation.

\subsubsection{Trajectory Logging}
\label{sec:trajectory-logging}

To support downstream safety evaluation, SeClaw records each sandbox execution as a structured trajectory covering the chain from task loading to final artifact generation. Each run begins with a task-specific configuration, including the scenario identifier, task instruction, skill description, operation card, fixture files, and permission rules. The Docker/OpenClaw runner loads this configuration, initializes the sandbox, starts the required mock services, and connects the agent through the MCP bridge. Thus, the trajectory captures not only the final output, but also intermediate model decisions, tool invocations, service responses, audit events, and generated artifacts.
Formally, for a task instance $i$, SeClaw represents the execution trajectory as
\begin{equation}
\tau_i =
\left(
c_i,
{ \ell_t }_{t=1}^{T_i},
y_i,
\omega_i
\right),
\end{equation}
where $c_i$ is the task configuration, $\ell_t$ is the interaction log at step $t$, $y_i$ is the final output, and $\omega_i$ denotes persisted artifacts such as LLM traces, task metadata, runtime logs, generated files, and integrity records. This treats the complete sandbox run as a first-class evaluation object rather than reducing agent behavior to a single final answer. Each step-level log links model-side decisions with tool- and service-level execution:
\begin{equation}
\ell_t =
\left(
p_t,
u_t,
a_t,
g_t,
v_t,
b_t
\right),
\end{equation}
where $p_t$ is the prompt or context, $u_t$ is the model response, $a_t$ is the agent action, $g_t$ is the MCP tool call, $v_t$ is the mock service response, and $b_t$ is the associated audit record. This allows SeClaw to reconstruct how the agent interprets the task, selects actions, invokes external capabilities, receives feedback, and continues execution.
The logs are organized into three views: the \emph{model trajectory}, which records prompts, responses, decisions, and tool-use intentions; the \emph{tool and service trajectory}, which records MCP calls, service inputs/outputs, timestamps, execution status, and observations; and the \emph{artifact trajectory}, which records files, logs, metadata, and final outputs. Together, these views provide evidence for identifying whether unsafe outcomes stem from model interpretation, action selection, tool invocation, or environment response, enabling fine-grained safety scoring, failure attribution, and reproducible behavioral analysis.

\subsubsection{Security Evaluation}
\label{sec:security-evaluation}

Given the structured trajectories produced by the logging module, SeClaw performs security evaluation by analyzing whether an agent's execution violates the safety constraints specified by each task. The evaluator takes as input the task configuration $c_i$ and the recorded trajectory $\tau_i$, and produces a sample-level judgment indicating whether the execution successfully triggers the intended safety-risk condition. In contrast to evaluations that consider only the final response, SeClaw evaluates both the outcome and the execution process, including tool invocations, service interactions, file operations, permission usage, and generated artifacts.

Formally, for each submitted sample $i$, the trajectory analyzer produces a binary sample-level outcome
\begin{equation}
    r_i = \mathcal{J}(c_i, \tau_i) \in \{0,1\},
\end{equation}
where $r_i=1$ indicates that the sample reaches the predefined unsafe condition under the task-specific evaluation rules, and $r_i=0$ otherwise. Each task is associated with a target identifier $o_i \in \mathcal{O}$, where $\mathcal{O}$ denotes the set of predefined security targets. This allows the scoring engine to aggregate sample-level outcomes into benchmark-level metrics that capture both breadth and reliability.

Let $N$ denote the total number of submitted samples, $S_N = \sum_{i=1}^{N} r_i$ the number of successful samples, $O = |\mathcal{O}|$ the number of predefined targets, and $S_O$ the number of unique targets successfully reached by at least one sample. We define the coverage score as
\begin{equation}
    C = \frac{S_O}{O}, \qquad C \in [0,1],
\end{equation}
which measures how broadly an agent or attack strategy covers distinct security targets. We further define the attack success score as
\begin{equation}
    P = \frac{S_N}{N}, \qquad P \in [0,1],
\end{equation}
which measures how reliably the submitted samples succeed across all evaluated instances.

To jointly account for both dimensions, SeClaw computes an overall attack score using the harmonic mean:
\begin{equation}
    F_{\mathrm{attack}} =
    \frac{2CP}{C+P},
    \qquad
    F_{\mathrm{attack}} \in [0,1].
\end{equation}
This formulation follows the intuition of the $F_1$ score: a high final score requires both broad target coverage and high sample-level reliability. As a result, strategies that repeatedly succeed on only a narrow subset of targets, as well as strategies that attempt many targets but succeed inconsistently, are both penalized. When either $C=0$ or $P=0$, we set $F_{\mathrm{attack}}=0$.

Importantly, $F_{\mathrm{attack}}$ should be interpreted as a \emph{risk-oriented} score rather than a utility-oriented performance score. A larger value indicates that the evaluated base model is more susceptible to the corresponding safety-risk tasks, since unsafe conditions can be triggered both broadly across targets and reliably across submitted samples. Therefore, a higher $F_{\mathrm{attack}}$ implies weaker safety robustness of the model under the evaluated task distribution, and also suggests that the associated safety-risk task or attack strategy poses a greater practical threat. Conversely, a lower score indicates that the model either resists most unsafe executions, limits them to a small subset of targets, or fails to complete the unsafe behavior consistently.





\section{Further Exploration}
\label{sec:further-exploration}

SeClaw is designed as an extensible benchmark, and we plan to expand it along several directions. First, we will release evaluation results on a set of foundation models, including Qwen, Kimi, GPT, Gemini, and other representative model families, to provide a comparative view of safety robustness under the same safety-risk task distribution. Second, SeClaw will be integrated into additional agent execution harnesses, such as Claude Code and other general-purpose coding or tool-using agents, in order to evaluate whether the observed risks persist across different agent infrastructures and interaction protocols. Third, we will further investigate implicit safety-risk injection during intra-agent task propagation, where unsafe objectives or constraints may be transformed, hidden, or amplified as tasks are passed across planning, memory, tool-use, and execution modules. Finally, we will continue improving the benchmark along task diversity, evaluator calibration, and trajectory-level interpretability, so that SeClaw can support more reliable measurement of agent safety in realistic, stateful execution environments.


\bibliography{iclr2025_conference}
\bibliographystyle{iclr2025_conference}

\appendix
\section{Appendix}
You may include other additional sections here.

\end{document}